\documentclass[11pt]{article}
\usepackage{amssymb,dsfont,amsmath,amsfonts,hyperref,mdframed,mathrsfs,stmaryrd,bbold}
\usepackage[active]{srcltx}
\usepackage{slashed}
\usepackage{color}
\advance \textheight by \topskip
\DeclareMathAlphabet{\mathpzc}{OT1}{pzc}{m}{it}
\usepackage{xcolor}
\definecolor{mygreen}{rgb}{0.1, 0.7, 0.2}

\def\be{\begin{equation}}
\def\ee{\end{equation}}
\def\bea{\begin{eqnarray}}
\def\eea{\end{eqnarray}}
\def\oh{h}

\def\a{\alpha}
\def\b{\beta}

\def\g{\gamma}

\def\bfPsi{\mathbf{\Psi}}
\def\Comp{\mathbb{C}}
\def\Real{\mathbb{R}}

\def\ie{\textit{i.e.} }
\def\eg{\textit{e.g.} }
\def\cf{\textit{cf.} }

\def\bos{\texttt{bos} }
\def\fer{\texttt{fer} }
\def\bterm{\hbox{b.t.}}
\def\BOS{ \vert_{\scriptstyle{\texttt{BOS}} }}
\def\FER{\vert_{\scriptstyle{\texttt{FER}} }}
\def\ST{ \vert_{\scriptstyle{\texttt{ST}} }}
\def\INT{\vert_{\scriptstyle{\texttt{INT}} }}
\def\LOR{\vert_{\scriptstyle{\texttt{L}} }}
\def\TRA{ \vert_{\scriptstyle{\texttt{T}} }}
\def\DIL{\vert_{\scriptstyle{\texttt{D}} }}
\def\eff{{}^{\scriptstyle{\texttt{eff}} }}
\def\one {\mathds{1}}
\newcommand{\JJ}{\mathds{B}}

\newcommand{\DD}{\mathds{D}}
\newcommand{\TT}{\mathds{B}}

\newcommand{\QQ}{\mathds{Q}}
\newcommand{\AAA}{\mathds{A}}
\newcommand{\FF}{\mathds{F}}
\newcommand{\WW}{\mathds{W}}
\newcommand{\UU}{\mathds{U}}
\newcommand{\II}{\mathds{I}}
\newcommand{\BB}{\mathds{B}}

\newcommand{\XX}{\mathds{X}}
\newcommand{\GG}{\mathds{G}}

\newcommand{\calL}{\mathcal{L}}

\newcommand{\calN}{\mathcal{N}}

\newcommand{\mfh}{\mathfrak{h}}
\newcommand{\mfg}{\mathfrak{g}}
\newcommand{\mff}{\mathfrak{f}}
\newcommand{\mfA}{\mathcal{A}}
\newcommand{\mfF}{\mathcal{F}}
\newcommand{\mfD}{\mathcal{D}}
\newcommand{\mbfM}{\mathbf{M}}

\newcommand{\calX}{\mathcal{X}}
\newcommand{\calK}{\mathcal{K}}
\newcommand{\str}{\texttt{\,str\,}}
\newcommand{\tr}{\texttt{\, tr\,}}
\newcommand{\cast}{\circledast}

\numberwithin{equation}{section}

\begin{document}

\title{$\calN=2$ Extended MacDowell-Mansouri Supergravity
}

\author{
Pedro D. Alvarez$^1$, Lucas Delage$^{2,3}$, Mauricio Valenzuela$^3$ and Jorge Zanelli$^3$\\[12pt]
$^1$ {\small \textit{Departamento de F\'{\i}sica, Universidad de Antofagasta, Aptdo. 02800, Antofagasta, Chile}}\\[5pt]
$^2$ {\small \textit{Instituto de Matem\'atica y F\'{\i}sica, Universidad de Talca, Casilla 747, Talca, Chile} 
}  \\[5pt]
$^3$ {\small  \textit{Centro de Estudios Cient\'{\i}ficos (CECs), Arturo Prat 514, Valdivia, Chile }}
}

\date{\today}

\maketitle

\begin{abstract}

We construct a gauge theory based in the supergroup $G=SU(2,2|2)$ that generalizes MacDowell-Mansouri supergravity. This is done introducing an extended notion of Hodge operator in the form of an outer automorphism of $su(2,2|2)$-valued 2-form tensors.
The model closely resembles a Yang-Mills theory---including the action principle, equations of motion and gauge transformations---which avoids the use of the otherwise complicated component formalism. The theory enjoys $H=SO(3,1)\times \mathbb{R} \times U(1)\times SU(2)$ off-shell symmetry whilst the broken symmetries $G/H$, translation-type symmetries and supersymmetry, can be recovered on surface of integrability conditions of the equations of motion, for which it suffices the Rarita-Schwinger equation and torsion-like constraints to hold. Using the \textit{matter ansatz}---projecting the $1 \otimes 1/2$ reducible representation into the spin-$1/2$ irreducible sector---we obtain (chiral) fermion models with gauge and gravity interactions.

\end{abstract} 


\newpage

\tableofcontents 

\newpage
%%%%%%%%%%%%%%%%%%%%%%%%%%%%%%%%%%%%%%%%%%%%%%%%%
\section{Introduction}
%%%%%%%%%%%%%%%%%%%%%%%%%%%%%%%%%%%%%%%%%%%%%%%%%

Yang-Mills theories have an important r\^ole in  physics. They are a paradigm of mathematically interesting objects which play an essential r\^ole in the standard model, accurately describing the electroweak and strong forces. The relevant force at large distance---gravity---is described by the Einstein-Hilbert action which represents a different paradigm of predictive field theory. 

Besides their apparent dissimilar structure, these field theories have another important difference. The fundamental fields of gravity---in the first order formulation---are the vielbein and the spin connection, associated to local translations and Lorentz transformations, respectively. However, only the Lorentz (structure) group is a gauge invariance of the Einstein-Hilbert action, while  local translation invariance is a broken symmetry \cite{Regge:1986nz}.  Lorentz invariance does not require additional conditions and is therefore called an \textit{off-shell} symmetry. In order to enforce local translation invariance of the action, it is necessary to appeal to the so-called \textit{torsion constraint}, which is a consequence of the field equations and it is therefore referred to as an \textit{on-shell} symmetry. In contrast, Yang-Mills actions are invariant with respect to the entire symmetry group, requiring no additional conditions on the dynamical fields. 

In \textit{pure supergravity} \cite{fnf,dz}---composed  by the Einstein-Hilbert and the Rarita-Schwinger actions---supersymmetry remains on-shell \cite{f-van} up to a torsion constraint, $T^a\cong \bar{\psi}\gamma^a \psi$, like in first order gravity. The introduction of auxiliary fields \cite{sw,fn} makes it possible to realize the off-shell fermionic symmetry (for further details see \eg \cite{Freedman:2012zz,RauschdeTraubenberg:2020kol,Ducrocq:2021vkh}).

Leaving aside the introduction of auxiliary fields, on-shell and off-shell symmetries  play different roles in (super)gravity. As is well known, off-shell symmetries can be represented by a principal bundle. Broken off-shell symmetries on the other hand, which are preserved when some constraints are imposed, could be understood as sections of an associated vector bundle. Indeed, as both on- and off-shell symmetries form a group, there is a natural representation of the structure group on the generators of infinitesimal on-shell symmetries.

Symmetries that are realized on the surface of the field equations and which belong to certain Lie (super)group that also contains (unbroken) off-shell symmetries are often referred to as ``on-shell symmetries".\footnote{The expression ``on-shell symmetry" appears to us as vacuous, since any transformation of a field $\delta\mfA$ leaves invariant the action on-shell. Instead of on- and off-shell symmetries, we prefer to refer to them as unconditional or conditional symmetries, respectively.} The equations of motion provide \textit{sufficient but not necessary} conditions for these symmetries to hold; the consistency conditions (\cf, $\tilde\Upsilon=0$ in section 2), provide \textit{necessary conditions} for the invariance of the action and they can be therefore called \textit{symmetry constraints}. The symmetry constraints are, in general, less restrictive than the equations of motion and we shall refer to the symmetries that arise when these constraints hold as \textit{conditional symmetries}.

The MacDowell-Mansouri approach \cite{MacDowell:1977jt} of pure SUGRA shows clearly this pattern. Their supergravity action principle is a quadratic form of the gauge curvature for a $osp(4|1)$-valued connection, however, this bilinear explicitly breaks the $OSp(4|1)$ symmetry leaving unbroken only the Lorentz subgroup. The translation symmetry is broken and the corresponding ``dual" symmetry constraint holds on the surface of the torsion constraint. Supersymmetry is also broken and the  dual symmetry constraint appears as a product of the torsion and the fermion curvature, which is therefore automatically satisfied also imposing the torsion constraint.

In this paper we examine a gauge theory that realizes once more these ideas, now for the \textit{extended supersymmetry} $SU(2,2|2)$. The bosonic subgroup is given by $SO(4,2)\times U(1)\times SU(2)$, associated to the spin connection, two transvection (\ie non-commutative translation) type fields, dilation, the electroweak-like gauge fields and $\calN=2$ supercharges, associated to two spin-$3/2$ gravitino fields. 

As we shall see, our model extends the MacDowell-Mansouri formulation of pure supergravity \cite{MacDowell:1977jt} and the Yang-Mills action for $U(1)\times SU(2)$ internal gauge fields. Thus both, gravity and $U(1)\times SU(2)$ interactions, are unified in a single action principle.  In contrast to the usual $\mathcal{N}=2$ supergravities \cite{deWit:1979dzm,deWit:1980lyi,deWit:1983xhu,deWit:1984wbb,deWit:1984rvr,Castellani:1990tp,DAuria:1990qxt}, where vector and matter multiplets are required, our model is built upon the gauge fields of the $su(2,2|2)$ connection only. In addition, instead of Majorana supercharges we consider Dirac supercharges that make the $R$-symmetry manifest.

An advantage of our approach is that we can make extensive use of  differential geometry techniques which allows a homogeneous treatment of pure supergravity and internal Yang-Mills theories. Computations of gauge transformations of the action or the obtention of the equations of motion are greatly simplified, resulting in compact expressions, closely resembling those in standard Yang-Mills theories. 

The paper is organized as follows: In section \ref{sec:summary} we present the general framework in which our work is situated and we summarize briefly the main result. In section \ref{sec:math} we introduce the notation. In section \ref{sec:GYM} we present the Lagrangian of our theory, find the field equations and symmetries. In section \ref{sec:effe} we explore the supersymmetric ground states and the perturbation theory around them. In \ref{sec:mansatz} we construct models for spin-$1/2$ fermion matter fields using the \textit{matter ansatz}. Section \ref{sec:conc} contains the conclusions.

%%%%%%%%%%%%%%%%%%%%%%%%%%%%%%%%%%%%%%%%%%%%%%%%%%%
\section{Conditional symmetries}\label{sec:summary}
%%%%%%%%%%%%%%%%%%%%%%%%%%%%%%%%%%%%%%%%%%%%%%%%%

In this section, we present the general scheme of our model without technical details. This theory will be built from a Lie (super)algebra-valued connection one-form that contains all the fields appearing in the action, as in a standard gauge theory. More precisely, let $\mfg$ be a super algebra, $\mfA \in \mfg$ the gauge potential, and $S_{\textit{pt}}[\mfA]$ the corresponding action principle.
The variation of the action with respect to the $\mfA$ reads,
\be\label{daproto}
\delta S_{\textit{pt}}=\int \; \delta \mfA^\mbfM \Upsilon_\mbfM\,+ \bterm,
\ee
where $\mbfM$  labels the (super)algebra generators and the differential operator $\Upsilon_\mbfM=\Upsilon_\mbfM[\mfA]$ is ``dual'' to $\delta\mfA$. The action is invariant under the proposed variation if $\Upsilon_\mbfM=0$, which defines the field equations of the system. In what follows we shall often omit the $\wedge$-product of differential forms, and assume that the (anti)commutator $[\,,\,]$ is graded with respect the form degree and statistics of the fields, consistently with the Lie (super)algebra under consideration.

For gauge transformations the transformation parameter takes the particular form $\delta \mfA^M = (\mfD \lambda)^\mbfM$, where $\mfD$ is the covariant derivative. Clearly the action remains invariant for ``Killing vectors" parameters $\lambda$, $\mfD\lambda=0$, or when $\lambda$ is in the kernel of $\tilde\Upsilon$. When this is not the case, upon partial integration \eqref{daproto} yields,
\be\label{dgproto}
\delta_\lambda S_{\textit{pt}}=\int \; \lambda^\mbfM \tilde\Upsilon_\mbfM\, + \bterm,
\ee
where the dual differential operator $\tilde\Upsilon_{\mbfM}\cong (\mfD \Upsilon[\mfA])_{\mbfM}$ is dual to the parameter $\lambda^\mbfM$. It turns out that $\tilde\Upsilon =0$ is an integrability condition for the equation of motion $\Upsilon=0$ and, at the same time, an indicator of whether the parameter $\lambda^\mbfM$ generates a symmetry or not.

Consistently with our previous definition, an \textit{off-shell symmetry} is the one for which $ \tilde \Upsilon_\mbfM \equiv 0$ is an identity. Reciprocally, a   \textit{conditional gauge symmetry} is one for which $\tilde \Upsilon_\mbfM$ does not vanish identically, but needs to be imposed as a constraint $\tilde \Upsilon_\mbfM = 0$. Thus, the index $\mbfM$ in \eqref{dgproto} can be restricted to run over the off-shell broken symmetry generators only.

Geometrically, the distinction between off-shell and conditional symmetries can be understood as follow. Let $\mathfrak{g}\ni \mfA$ be the Lie (super)algebra generating the a Lie (super)group $G$---which combines off-shell and  conditional symmetries---and let $\mfh$ be the algebra of off-shell symmetries generating the subgroup $H\, \subset \, G$. The broken gauge symmetries are those in the coset $G/H$, spanned locally by the vector subspace $\mff \, \subset\, \mfg$. Denoting by $\mathfrak{h}$ the Lie subalgebra corresponding to the subgroup H, we can decompose $\mathfrak{g} = \mathfrak{h} \oplus \mathfrak{f}$, where the direct sum is a vector space direct sum. The differential operators $\tilde \Upsilon_{\mbfM}$ dual to the generators of $\mfh$ vanish identically whilst those dual to the generators of $\mff$ do not. 

Since $[\mfh\,,\mff]$ is a subset of $\mff$ the gauge fields valued in the algebra elements $\mff$ transform as vectors under the endomorphisms generated by the unbroken symmetry algebra $\mfh$.  Effectively, the gauge fields components in $\mff$ can be regarded as fiber bundle sections. The group $H$ is the real structure group of the fiber bundle. From this perspective, the group $G$ puts together the structure group $H$ and fiber sections, unifying the $H$-principal bundle and the associated fiber bundle with sections in $\mff$. For a more rigorous exposition of these subjects see \eg \cite{Eder:2020erq,Eder:2021nyb}.

In order to illustrate these aspects we shall consider briefly Yang-Mills theories and the MacDowell-Mansouri approach \cite{MacDowell:1977jt} for $\calN=1$ (super)gravity. In the first case, the  action reads, 
\begin{equation}\label{YM}
S_{\texttt{YM}}\;=  \int \tr\; \mfF \, * \mfF\;,
\end{equation}
where $\mfF=d\mfA+\mfA^2$ is the 2-form field strength of the gauge connection one form $\mfA\,\in\,\mfg$ and $*$ is the Hodge dual operator. 
The variation of the gauge field $\delta \mfA$ and the gauge transformation  $\delta \mfA=\mfD\lambda$ yields respectively \eqref{daproto} and \eqref{dgproto} with dual differential operators,
\be\label{duals}
\Upsilon = \,D*\mfF \,,\qquad  \tilde \Upsilon  = \,[\mfF\,,\,*\mfF]\,.
\ee
For general Yang-Mills theories $[\mfF\,,\,* \mfF] \equiv 0$ and the whole symmetry group $G$ is preserved off-shell.

Next, the MacDowell-Mansouri action is given by 
\bea
S_{\texttt{MM}} &=& \int \; \mfF^\mathbf{M} \, Q_{ \mathbf{M N} }\, \mfF^\mathbf{N}\,,\label{MM1}
\eea
where now the field strength $\mfF$ is valued in the algebra $\mfg=so(3,2)$ for first order gravity, or in $\mfg=osp(4|1)$ for pure supergravity. Appealing to the standard nomenclature, in the first case, $G = SO(3,2)$ and $\mfA=\frac{1}{2}\omega^{ab}\JJ_{ab}+ e^a \JJ_{a}$; in the second case, $G=OSp(4|1)$ and $\mfA=\frac{1}{2}\omega^{ab} \, \JJ_{ab}+\, e^a \, \JJ_a +\, \overline{\QQ}_\alpha \, \psi^\alpha$. In both cases $Q_{\mathbf{M N}}$ is a Lorentz invariant tensor but it breaks explicitly the transvection symmetry generated by $\JJ_a$, and it also breaks the supersymmetry transformations generated by $\overline{\QQ}$. It can be shown that for the Lorentz transformation parameters the dual constraint vanishes identically, whilst for supersymmetry the dual constraints (see \eg \cite{vanNieuwenhuizen:2004rh}), 
\be
\tilde{\Upsilon}\; \cong \; \mfF^a \, \gamma_a\, D\psi=0\,,\label{susyMM}
\ee
where  $\mfF^a$ is the transvection-valued component of the field strength, $\g$ is a Dirac matrix and $D\psi$ is the covariant derivative for the $so(3,2)$ connection acting on the gravitino field in the Majorana representation. Although \eqref{susyMM} admits different solutions, it is often solved using the more restrictive field equations. Indeed, the equation obtained by extremizing the action with respect to the spin connection yields,
\be
{\Upsilon}_{cd} \cong \epsilon_{abcd} \; \mfF^a \, e^b =0\,, \label{EoML}
\ee
implying, for invertible vierbeins ($e^b_\mu$), the \textit{torsion constraint} 
\be\label{tconst}
\mfF^a=0\,.
\ee
Here $\mfF^a$ consists of the vierbein torsion in pure gravity, and in the case of supergravity it also contains an additional 2-form fermion-current. Hence  on the surface of the constraint \eqref{EoML} the action is invariant under local supersymmetry transformations. Alternatively, when the gravitino field strength $D\psi$ vanishes, or when it is in the kernel of $\mfF\gamma$, supersymmetry also holds. 

The constraint dual to the transvection transformation parameters is 
\be
\tilde\Upsilon \cong  \frac12  \epsilon_{abcd}\; \mfF^{bc} \;  \mfF ^d - \, D\overline{\psi}  \, \g_a  \,  i\g_5 D\psi =0 \,,\label{tUpsi}
\ee
which, using the torsion constraint \eqref{tconst}, reduces to
\be\label{aux}
D\overline{\psi}  \,  \g_a  \,  i\g_5 D\psi=0\,.
\ee
This constraint can be satisfied for example if 
\begin{equation} \label{ig5Dpsi}
i\g_5 D\psi =  \varphi D\psi + \varphi' * D\psi , 
\end{equation}
where $\varphi$ and $\varphi'$ are scalar fields. This is because $(C\g^a)_{\alpha \beta}$ is symmetric in its spinor indices, where $C$ is the conjugation matrix, whilst the products $(D\psi)^\a  (D\psi)^\b$ and  $(D\psi)^\a  *(D\psi)^\b $ are antisymmetric. Conditions \eqref{ig5Dpsi} can be fulfilled by configurations that are not necessarily solutions of the field equations, but it can be checked that on-shell configurations do satisfy the constraint \eqref{aux}. Indeed, the Rarita-Schwinger equation obtained by varying with respect to the gravitino,
\be\label{RS}
\slashed{e}\, D\psi=0\,,\qquad \mbox{where}\;\; \slashed{e}:=\g_a e^a,
\ee
implies that \eqref{aux} holds. In order to prove this, we can use the equivalent form of the Rarita-Schwinger equation in four dimensions (see \eqref{RSequiv}),
\be
(i\g_5-*)D \psi=0\,,
\ee
which is in the class of \eqref{ig5Dpsi} for $\varphi=0$ and $\varphi'=1$. Hence, both local supersymmetry and transvection invariance are conditional symmetries of $\calN=1$ supergravity.

In this article we consider an $\calN=2$ supergravity model following the same pattern, unifying the MacDowell-Mansouri supergravity and the non-abelian $U(1)\times SU(2)$ Yang-Mills theory. Our action principle can be expressed in the Yang-Mills fashion,
\be \label{action}
S :=- \int \str \;\; \mfF \circledast \mfF\; ,
\ee
(\cf \cite{Wise:2006sm} in the pure gravity case) where $\mfF=d\mfA+\mfA^2\;\in\; \mfg=su(2,2|2)$, $\str$ is the supertrace and $\circledast$ combines the standard Hodge operator and an involution of the superalgebra $su(2,2|2)$. The explicit form of $\circledast$ is given in Eq. \eqref{castdef}. 

The field equations and corresponding consistency conditions that follow from \eqref{action} can be written as
\bea\label{EoMintro}
\Upsilon \; \cong \;\mfD \cast {\mfF^+}=0\,,\qquad
  \tilde \Upsilon = \mfD ^2 \cast {\mfF^+}=[\mfF\,,\, \cast {\mfF^+}]=0\,,
\eea
where $\mfD$ is the $su(2,2|2)$ covariant derivative and ${\mfF^+}$ is the $su(2,2|2)$ curvature with the terms along transvection generators removed. The removal of the transvection terms is prompted by the $\cast$ operator, which is necessary in order to recover the pure (super)gravity sector.

The gauge transformation of the action, \eqref{dgproto}, takes the form
\be \label{varact}
\delta S := -\int \str \;\; \lambda\,  [\mfF ,\circledast \mfF^+]\; + \bterm\,,
\ee
hence the $su(2,2|2)$ symmetry holds on the surface of the non-trivial components of the integrability condition $\tilde\Upsilon$ in \eqref{EoMintro}. As we shall see $ [\mfF ,\circledast \mfF^+]$ vanishes trivially except for the terms along transvection generators and supercharges, analogously to $\calN=1$ supergravity. Thus, the group $G=SU(2,2|2)$ breaks into $H=SO(3,1)\times \Real \times U(1)\times SU(2)$ off-shell symmetries while transvections and supersymmetry are conditional symmetries.

In what follows we present these results in detail.

%%%%%%%%%%%%%%%%%%%%%%%%%%%%%%%%%%%%%%%%%%%%%%%%%
\section{Mathematical setup}  \label{sec:math}
%%%%%%%%%%%%%%%%%%%%%%%%%%%%%%%%%%%%%%%%%%%%%%%%%

Here we introduce the $su(2,2|2)$ superalgebra, the gauge connection and the field strength.  

%%%%%%%%%%%%%%%%%%%%%%%%%%%%%%%%%%%%%%%%%%%%%%%%%
\subsection{Superalgebra representation}
%%%%%%%%%%%%%%%%%%%%%%%%%%%%%%%%%%%%%%%%%%%%%%%%%

In this section, we consider $su(2,2|2)$ as spanned by
\bea
&\{\underbrace{\JJ_{ab}\,, \JJ_{a}\,, \tilde\JJ_{a} \,,\JJ_5}_{ so(4,2)}\, ; \:   \underbrace{\TT_6\,, \TT_{I}}_{u(1)\oplus su(2)} \,; \: \underbrace{\QQ_i^\alpha, \;\overline{\QQ}^i_\alpha}_{\hbox{supercharges}}\}\,. &\label{su222} 
\eea
This representation will allow us to handle complex gravitino fields charged under $U(1)\times SU(2)$ interactions (see the appendix \ref{app:su222} for more details).

Here the $so(4,2)$ generators are labeled by spacetime indices ($a,b$) in the range $0,1,2,3$, $su(2)$ indices ($I$) in the range $7,8,9$, and spinorial labels ($\alpha$) in the range $1,...,4$, whilst there is single $u(1)$ generator with the label $6$. Hence the whole set of internal symmetry generators are labeled by Latin letters, $r,s,...,$ in the range $6,7,8,9$, $\TT_r \,\in \,u(1)\oplus su(2)$.  The supercharges \textit{isospin} labels $i=1,2$, transform in the fundamental representation of $su(2)\oplus u(1)$. 

The adjoint action of the bosonic generators, denoted $\BB$, onto the fermionic generators
\begin{eqnarray} 
&[\QQ^\alpha_i,\BB_M]=(B_M)^{\alpha\, j}_{i\,\beta }\,\,\QQ^\beta_j, \qquad [\BB_M,\overline{\QQ}^i_\alpha]=\overline{\QQ}^j_\beta\,\, (B_M)^{\beta \,i}_{j\, \alpha}\,,&\label{com}
\end{eqnarray}
 provides  the fundamental representations of the spacetime symmetry algebra  $ so(4,2)$ and the internal symmetry algebra $u(1)\oplus su(2)$, by means of the structure constants $B_M$, where the indices $M\in \{a,[ab],r\}$ label Lorentz vectors/tensors and internal symmetry generators. Other (anti-)commutation relations can be found in \eqref{BB1}-\eqref{anticom0}.

The constant of structures $B_M$ can be expressed in terms of tensor products involving $4\times4$ spinor representations for spacetime symmetry generators,
\be 
 so(4,2): \qquad (B_M)^{\alpha \, j}_{i\,\beta} = \delta_i{}^j \,\times\, \Big\{\frac12 (\g_{ab})^\a{}_\b \, , \quad \frac12 (\g_{a})^\a{}_\b \,,\quad \frac12 (\g_{a}\g_5)^\a{}_\b \,,\quad \frac12 (\g_5)^\a{}_\b \,\Big\}\,,\label{rhoJJ}
\ee
or from $2\times 2$ matrices for internal symmetries 
\be
u(1) \oplus su(2): \qquad (B_M)^{\alpha \, j} _{i\, \beta} = \delta^\alpha{}_\beta \times \Big\{  -i(\mathds{1}_{2\times2})_i{}^j\,, \quad -\frac{i}{2}(\sigma_I)_i{}^j \Big\} \,, \label{rhoTT}
\ee
where $\g$'s are Dirac gamma matrices and $\sigma_I$ are the Pauli matrices \eqref{pauli}. We shall denote the adjoint representation
\be\label{rho}
\rho(\BB_M)=B_M\,,
\ee 
simply by $\rho$-representation.

Introducing the Killing form $\calK_{MN}$ normalized by,
\be
\str (\BB_M \BB_N) =\calK_{MN}\,,\label{strB}
\ee
the anti-commutator $[\, \QQ , \overline{\QQ} \,]_+$ can be cast in a compact form using the representation $\rho$ of the bosonic subalgebra and the inverse Killing form $\calK^{MN}$,
\be \label{anticom1}
[\,\QQ,\overline{\QQ}\,]_+=\calK^{MN} \, B_M \;  \BB_N\,,
\ee 
where the $B_M$s are given in \eqref{rhoJJ}-\eqref{rhoTT}. Note that the $su(2,2|2)$ contains the subalgebras, 
\bea
so(3,2)&=& \{\JJ_{ab}\,,\JJ_{a}\} \; \cong \, sp(4)\,,  \label{ads}
\\[4pt]
so(4,1) &=& \{\JJ_{ab}\,,\tilde\JJ_{a}\}\,,\label{ds}
\\[4pt]
iso(3,1)_{\pm} &=&\{\JJ_{ab}\,, \BB_{\pm\, a}\}\,, \qquad \BB_{\pm\, a}:= \frac{1}{2}(\JJ_a \pm \tilde\JJ_{a})\,,\label{poinc}
\eea
which are isometries correspondingly of anti-de Sitter, de Sitter and Minkowski spacetimes. In the latter case we have two options, $iso(3,1)_+$ or $iso(3,1)_-$, for Poincar\'e subalgebras.

%%%%%%%%%%%%%%%%%%%%%%%%%%%%%%%%%%%%%%%%%%%%%%%%%
\subsection{The gauge potential}
%%%%%%%%%%%%%%%%%%%%%%%%%%%%%%%%%%%%%%%%%%%%%%%%%

The gauge potential (connection) is of a one-form valued in the superalgebra \eqref{su222},
\bea
\mfA:&=&\AAA +\bfPsi - \overline{\bfPsi} \quad \in \quad su(2,2|2) \,,\label{fkA}\\[5pt] 
\AAA:&=& A^M \BB_M\,, 
\eea
which we have decomposed in its fermion sector containing the gravitino supercharge-valued field, $\bfPsi:=\psi_i^\a \, \overline{\QQ}_\a^i$ and its conjugate $\overline{\bfPsi} := \overline{\psi}_\a^i \,\QQ_i^\a$, and the bosonic sector containing spacetime ($\WW$) and internal ($\UU$) symmetry components, 
\bea
&\AAA= \mathds{W}+\mathds{U}\,,&\label{AAA}\\[5pt]
&{\WW}=\frac12 \omega^{ab}\JJ_{ab}+p^a\, \JJ_{a} +\tilde{p}\, {}^a\, \tilde\JJ_{a} +h\, \JJ_5\qquad \in \quad so(4,2)\,,&\label{WW}\\[5pt] 
&\UU= U^r \TT_r \qquad \in \quad u(1)\oplus su(2)\,.&\label{UU}
\eea
Here $\omega^{ab}$ is the Lorentz connection, $p^a$ and $\tilde p\, {}^a$ are respectively $AdS_4-$ and $dS_4-$ type transvection gauge fields (\cf respectively \eqref{ads} and \eqref{ds}), $h$ is the dilation gauge field, $U^6$ is the $u(1)$ \textit{electromagnetic} gauge field and $U^I$ are $SU(2)$ gauge fields.

Using the adjoint representation $\rho$ \eqref{rho}, which does not affect the field coefficients, we can map the bosonic gauge connection $\AAA$ to its adjoint action $A:=\rho(\AAA)$ upon the gravitino fields,
\bea
&A=W+U \,,\qquad {W}=\rho(\WW)\,,\qquad U=\rho(\UU)\,,& \nonumber\\[5pt]
&{W}=\Omega+P+\tilde{P}+H\,,\qquad U=U^r  B_r\,,& \label{rhoA}
\eea
where 
\be\label{Acomp}
\Omega=\frac12 \omega^{ab}B_{ab}\,,\quad P=p^a B_{a}\,,\quad \tilde{P}=\tilde{p}\, {}^a \tilde B_{a}\,,\quad H=\oh B_5,\quad U^r B_r=U^6 B_6 +U^I B_I\,.
\ee
%%%%%%%%%%%%%%%%%%%%%%%%%%%%%%%%%%%%%%%%%%
\subsection{The field strength}
%%%%%%%%%%%%%%%%%%%%%%%%%%%%%%%%%%%%%%%%%%

The  covariant  derivative associated to the gauge connection \eqref{fkA} acts on $su(2,2|2)$-valued differential forms as
\be\label{covdev}
\mfD \, \Phi:=d\Phi\,+\,[\mfA\,,\, \Phi]\,.
\ee
We also introduce the covariant derivative, with respect to the bosonic gauge connection \eqref{AAA},
\be\label{covdevbos}
\DD:=d\,+\, \AAA\,.
\ee
The $su(2,2|2)$ field strength, $\mfF :=d\mfA+\mfA  \mfA$, has components 
\be\label{auxF}
\mfF =\frac12  \mfF^{ab} \JJ_{ab} +   \mfF\,{}^a \JJ_{a} +   \tilde \mfF\,{}^a \tilde \JJ_{a} +   \mfF^5 \JJ_5 +  \mfF^r \TT_r + \overline{\QQ}_\a^i\:\calX^\a_i -  \overline{\calX}_\a^i\, \QQ^\a_i \,, 
\ee
where
\bea
&\calX^\a_i = (D\psi)^\a_i\,,\quad \overline{\calX}_\a^i=(D \overline{\psi})_\a^i\,,& \label{auxX} \\[4pt]
&D{\psi} = d{\psi}+(W+U)\psi\,,\qquad D\overline{\psi} = d\overline{\psi}+\overline{\psi}(W+U)&
\eea
The covariant derivative $D= d+W+U$ is induced by the action of \eqref{covdevbos} on supercharge-valued gauge fields $\XX=\DD \bfPsi - \DD\overline{\bfPsi}$, where $\DD\bfPsi=\overline{\QQ}D\psi$ and $\;\;\DD\overline{\bfPsi}=(D\overline{\psi})\,\QQ$.

We identify three main sectors of the gauge curvature:
\bea
& \mfF = \FF-\II +\XX\,,\label{curvature}&\\[5pt]
&\FF:=\DD^2=F^{M}\, \BB_M\,,\qquad \II:=[\overline{\bfPsi}\,, \bfPsi]= I^M \BB_M\,\,,\qquad \XX:= \overline{\QQ}\:\calX -  \overline{\calX}\QQ\,, & \label{FFXX} 
\eea
respectively the bosonic gauge field strength, the gravitino (bosonic) 2-form current, where
\begin{eqnarray}  
&I^M=\calK^{MN}\overline{\psi}B_N\psi,& \\[4pt]
&I^{ab}=- \frac{1}{2}\overline{\psi}B^{ab}\psi
 \,,&\label{Ilorentz}\\[4pt]
&I ^a=\overline{\psi}B^a\psi \,,\qquad  \tilde I^a= - \overline{\psi}\tilde B^a\psi  \,,& \label{Itransv}\\[4pt]
&I^5=\overline{\psi}B_{5}\psi \,,&\label{Idil}\\[4pt]
&I^6=\frac14 \overline{\psi}B_6\psi \,,\qquad I^I=2 \overline{\psi}B_I\psi \,.\label{Iint}&
\end{eqnarray}

Thus, the boson and fermion components of the curvature are respectively,
\be
\mfF \BOS = \mfF^M \BB_M = \FF-\II\,,\qquad \mfF\FER=\XX\,.
\ee
For future reference, we shall use the ``evaluate'' symbol to project the superalgebra-valued differential forms on particular elements of the algebra, namely, $\BOS$ and $\FER$ to be the projections onto the bosonic and the fermionic sectors,  $\ST$ and $\INT$ the projections onto the spacetime and internal generators, $\LOR$, $\TRA$ and $\DIL$ the projections onto Lorentz, transvection and dilation generators.

Thus the boson components of $\mfF\BOS$ can be subdivided in their spacetime and internal type of components, $\FF\ST$ and $\FF\INT$, respectively given by,
 \be
 \FF := \FF\ST+\FF\INT \,, \qquad \FF\ST = d \WW + \WW \WW \,,  \quad \FF\INT =d\UU+\UU\UU 
\,.\label{FFsplit}
\ee
The 2-form current $\II= \II\ST+\II\INT$ is decomposed similarly. 

In more detail we have,
\be
\FF\LOR=\frac12 F^{ab} \JJ_{ab}\,,\quad  
\FF\TRA= F^a \JJ_{a} + \tilde F^a  \tilde\JJ_{a}\,,\quad
\FF\DIL=F^5 \, \JJ_5 \,,\qquad
\FF\INT=  G^6 \TT_6+ G^I \TT_I\,,\label{Fsplit}
\ee
where
\begin{eqnarray}
 &F^{ab} = R^{ab}(w)+p^ap^b-\tilde{p}^a\tilde{p}^b \,,\qquad R^{ab}(w):=dw^{ab}+w^{ac}w_c{}^{b}\,,&\label{Florentz}\\[4pt]
 &F^a = D_\Omega p^a-\oh \tilde{p}^a 
 \,,\qquad \tilde{F}^a = D_\Omega \tilde{p}^a -\oh {p}^a \,,&
 \label{Ftransv}\\[4pt]
&F^5 = d\oh+ p^a \tilde{p}_{a} 
\,,&\label{Fdil}\\[4pt]
&  G^6=dU^6\,,\qquad  G^I=dU^I+U^JU^K\epsilon_{JK}{}^I\,,& \label{FYM}
\end{eqnarray}
and for the gravitino currents,
 \begin{eqnarray}  
&\II\LOR=\frac12\,  I^{ab}\; \JJ_{ab}\,,\quad 
\II\TRA= I^a\; \JJ_a + \tilde I^a \;\tilde \JJ_{a} \,,\quad
\II\DIL=I^5 \; \JJ_5 \,,\qquad \II\INT = I^6\,\TT_6 + I^I\,\TT_I\,.&\label{Isplit}
\end{eqnarray}

In the adjoint representation \eqref{rho} we write, from \eqref{Fsplit}
\be
F:=\rho(\FF)= F^M B_M\,, \qquad I:=\rho (\II)= I^M B_M \label{FI}
\ee
In what follows we shall also use gauge-field symbols as labels in the covariant derivative in order to specify the gauge connection being used, 
$$
D_W= d+W\,,\quad D_U= d+U\,,\quad D_\Omega=d+\Omega\,,\quad D_{\Omega+H}=d+\Omega+H \,.
$$

\subsection{$\Gamma$--gradding}

The matrix 
\be
\Gamma = \left[\begin{array}{c|c}
\gamma_5 &  0_{4\times2}\\[0.5em] \hline
0_{2\times4} & 0_{2\times2} \\
\end{array}\right]\,.\label{Gamma}
\ee
induces a natural graded structure on the bosonic generators of $su(2,2|2)$, 
\be\label{grading}
[\BB^-_M,\Gamma]_+=0\,,\qquad [\BB^+_M,\Gamma]=0\,, 
\ee 
where $\BB^-={\JJ}_a\,,\tilde{\JJ}_a$,  and $\BB^+={\JJ}_{ab}\,,\tilde{\JJ}_5\,, \TT_r\,$.

The grading \eqref{grading} of the bosonic component of any differential form $\Theta\in\mfg$ is preserved in the representation $\rho$. Since $\rho(\Gamma)=\g_5$, we have $[\, \rho(\Theta\BOS^-) \, ,\gamma_5]_+=0$, $[\, \rho(\Theta\BOS^+) \, ,\gamma_5]=0$. Thus we can also decompose the differential forms valued in fundamental representations of the bosonic gauge algebra accordingly.  In particular, for future reference, the gauge connection decomposition reads,
\be \label{A-A+}
A^-=W^-=P+\tilde P\,,\qquad A^+=W^++U=\Omega+H+U\,.
\ee
Henceforth all differential form $\Theta\in\mfg$ can be decomposed as follows,
\bea\label{Omegadec}
 &\Theta= \Theta^+ + \Theta^- \,,& \\ 
 &\Theta^-= \Theta\TRA = \Theta ^a \JJ_a + \tilde\Theta^a \tilde{\JJ}_a  \,,\quad \Theta^+= \Theta\BOS^+ + \Theta\FER \,,\quad\Theta\BOS^+= \Theta\LOR+\Theta\DIL+\Theta\INT\,.&\nonumber
 \eea
Hence the denoted ${}^+$--component contains all the generators of the corresponding gauge algebra \underline{excluding} transvection generators. The ${}^-$--components refer therefore only to the transvection components.
In particular, the transvection term of the field strength is given by, 
\be\label{F-}
\mfF^-=(F^a-I^a) \, \JJ_a + (\tilde F^a-\tilde I^a)  \, \tilde\JJ_a\,,\qquad F^-=(F^a-I^a) \, B_a + (\tilde F^a-\tilde I^a)  \, \tilde B_a\,.
\ee  

%%%%%%%%%%%%%%%%%%%%%%%%%%%%%%%%%%%%%%%%%%%%%%%%%%%%
\section{Lagrangian, dynamics and symmetries}\label{sec:GYM}
%%%%%%%%%%%%%%%%%%%%%%%%%%%%%%%%%%%%%%%%%%%%%%%%%%%%

With the necessary ingredients at hand, we can proceed with the generalized Yang-Mills action \eqref{action}. The corresponding Lagrangian density, built from the field strength $\mfF$  \eqref{curvature}, is given by,
\be \label{Lagrangian}
\calL := -\; \str\Big(\mfF \circledast \mfF \Big)\,.
\ee
In order to use standard deferential geometry techniques, as in Yang-Mills gauge theories, $\cast$ must produce an authomorphism of the complexified $su(2,2|2)$ two-form curvature: $\cast\mfF  \, \in \, sl(4|2,\Comp)$. In order to operate similarly to the regular Hodge dual in Lorentzian signature, we shall also require $\circledast^2=-\mathds{1}$ on the 2-forms. Although there are several options, we shall choose here the following action of $\cast$:
\be\label{dualFconf}
\begin{split}
\circledast \mfF =&
\frac12 (\cast \mfF^{ab}) \JJ_{ab} + (\cast \mfF^a) \JJ_{a} + \cast (\tilde \mfF^a ) \tilde \JJ_{a} + (\ast \mfF^5) \JJ_5  \\ 
& + (* \mfF^r) \TT_r +  \overline{\QQ}\:\cast\calX -  (\cast\overline{\calX})\, \QQ  \,, 
\end{split}
\ee
where
\be\label{castdef}
\begin{gathered}
\cast \mfF^{ab}=  \frac12 \epsilon^{ab}{}_{cd} \mfF^{cd}\,,\quad \cast \mfF^a= -i  \tilde \mfF^a \,,\quad \cast\tilde \mfF^a= -i \mfF^a \,,\quad  \cast \mfF^5 = *  \mfF^5\,,  \\  
\cast \mfF^r = *  \mfF^r \,,\quad \cast\calX = i\gamma_5 \calX  \,,\quad \cast\overline{\calX} = \overline{\calX} i\gamma_5\,.
\end{gathered}
\ee
Variations of the signs of the $\cast$ operator on particular sectors of the curvature that may lead to different models. We shall discuss briefly two additional cases in sections \ref{sec:altcast} and \ref{sec:mansatz}. For more details see  Appendix \ref{sec:Lconst}.

Collecting only bosonic terms in $\calL_\bos$ and fermion terms in
$\calL_\fer$, is composed as,
\be
\calL = \calL_\bos \;+\calL_\fer
\ee
\be\label{Lfer1}
\calL_\fer=  4 \overline{\psi} \left( i\g_5 W^-D +  \frac12 (* - i \g_5)(F\DIL+F\INT) - i \g_5 \frac12 F^- - \frac14 \circledast {I^+} \right) \psi\,.
\ee
In components the bosonic component of the Lagrangian reads,
\be\label{Lbos1}
\begin{split}
\calL_\bos:=&\frac12 R^{ab}(w)p^{cd}\epsilon_{abcd} + \frac{1}{4} p^{ab}p^{cd}\epsilon_{abcd} + \frac{1}{4} R^{ab}(w)R^{cd}(w)\epsilon_{abcd} \\[4pt]
&- d\oh*d\oh - 2 d\oh*p^a\tilde{p}_{a} - p^a\tilde{p}_{a} \ast p^b\tilde{p}_{b} - \frac12 F^I*F^I - 4dU^0*dU^0\,,
\end{split}
\ee
where 
\be\label{pab}
p^{ab}:=p^a p^b -\tilde p\, {}^a\tilde p\, {}^b\,.
\ee

The Lagrangian contains boson and fermion kinetic terms at exception of the transvection gauge fields, couplings of fermion-currents and field strengths (Pauli couplings) and four-fermion self-interactions. 
It can be shown that the terms containing the transvection-like component of the curvature \eqref{F-} cancel out from the Lagrangian \eqref{Lagrangian} as a consequence of the $\cast$ action \eqref{castdef} along these terms. Hence the absence of  $F\TRA$ Pauli couplings and Kinetic terms is natural. The absence of $F\LOR$ Pauli couplings in  \eqref{Lfer1} is consequence of a cancelation of the identical terms provided by the boson gauge curvatures and the fermion gauge curvatures.

%%%%%%%%%%%%%%%%%%%%%%%%%%%%%%%%
\subsection{Field equations}
%%%%%%%%%%%%%%%%%%%%%%%%%%%%%%%%

The equations of motion are given by the vanishing condition of the variation of the action with respect to the gauge connection $\mfA$,  
\be
\delta \calL= - 2\,\str \left(\delta\mfA\;\mfD \cast \mfF^+ \right)  - \str\, d (\delta \mfA \cast \mfF ^+ )\,,\label{varL}
\ee
where ${\mfF}^+= \FF^+ -\II^+ + \XX$ from definition \eqref{Omegadec}.
 
From \eqref{varL} the equations of motion reads
\be\label{GYMeq1}
\mfD \cast {\mfF}^+=0\,.
\ee

In an extended form the equations of motion \eqref{GYMeq1} are given by:
\begin{itemize}
\item[$\delta w:$]
\be\label{EoMw1}
\epsilon_{cdab}\left[ (D_\Omega p^a-I^a\:) p^{b} - (D_\Omega\tilde p^a-\tilde{I}^a\:)  \tilde{p}^{ b}\right]=0\,.
\ee
This equation is equivalent to
\be\label{EoMw2}
\epsilon_{cdab}\left[ (F^a-I^a\:) p^{b} - (\tilde F^a-\tilde{I}^a\:)  \tilde{p}^{ b}\right]=0\,,
\ee
since the components $\oh p$ and $\oh \tilde{p}$ in the definitions \eqref{Ftransv} cancel. 

\item[$\delta h:$] 
\bea
d \left(\ast  F^5 -\frac{i}2\overline{\psi}\psi \right)=0\,.\label{EoMh}
\eea 

\item[$\delta p:$] 
\be
\frac12 \epsilon_{abcd} p^b (F^{cd}-I^{cd})-\tilde p _a \ast (F^5-I^5) -D\overline\psi  i \g_5 B_a \psi +\overline\psi B_a i \g_5 D\psi =0\,.\label{deltap}
\ee

\item[$\delta \tilde p:$]
\be
\frac12 \epsilon_{abcd} \tilde p^b (F^{cd}-I^{cd})- p _a \ast (F^5-I^5) +D\overline\psi  i \g_5 \tilde B_a \psi-\overline\psi \tilde B_a i \g_5 D\psi =0\,. \label{deltapt}
\ee

\item[$\delta U^r:$]
\be
 \DD_\UU \, \left(* \FF\INT + \II'\INT- * \II\INT \right)=0\,,\label{deltaUr}
\ee
where $ \II'\INT:=\overline{\psi}i\g_5B^r \psi \, \TT_r =\frac14 \overline{\psi}i\g_5B_6\psi \, \TT_6 + 2 \overline{\psi}i\g_5 B_I\psi \, \TT_I$.

\item[$\delta  \overline{\psi}:$]
\be
\left(i\g_5 W^- D + \frac12 (\circledast - i \g_5)(F^+-{I^+}) - \frac12 i \g_5  (F^--{I^-}) - \frac12 i \g_5 I  \right) \psi=0\,, \label{deltabpsi}
\ee
or alternatively,
\be\label{deltabpsi2}
\begin{split}
 \Big( i\g_5  W^- D_{\Omega+U} + i\g_5 (W^-)^2 + \frac12 (* - i \g_5)&((F-I)\DIL+(F-I)\INT)  \\ 
 -&\frac12 i \g_5  (D_\Omega W^- - {I^-})  -\frac12 i \g_5{I} \Big)\psi=0\,,
\end{split}
\ee
where $I$ is given in \eqref{FI}.

\item[$\delta  {\psi}:$]
Similarly,
\be
 D\overline{\psi} W^- \, i\g_5 +  \frac12 \overline{\psi} (\circledast - i \g_5)(F^+-{I^+}) - \frac12  \overline{\psi} (F^--{I^-})\, i \g_5 - \frac12 \overline{\psi} Ii\g_5 =0\,,\label{deltapsi}
 \ee
or alternatively,
\be  \label{deltapsi2}
\begin{split} 
(D_{\Omega+U} \overline{\psi}) W^-  i\g_5 + \overline{\psi}P^2 i\g_5  + \frac12 \overline{\psi}(* - i \g_5)&\Big((F-I)\DIL+(F-I)\INT\Big) \\ 
&-\frac12 \overline{\psi}(D_\Omega W^--{I^-})i \g_5    - \frac12 \overline{\psi}I i\g_5=0\,.
\end{split}
\ee

\end{itemize}

In \eqref{deltabpsi2} and \eqref{deltapsi2} we observe that the terms including the gauge field $H$ in the covariant derivative $D\Psi$ and in  $F^-=dW^-+[\Omega+H,W^-]$ cancel each other.

%%%%%%%%%%%%%%%%%%%%%%%%%%%%%%%%
\subsection{Integrability conditions and conditional symmetries}
%%%%%%%%%%%%%%%%%%%%%%%%%%%%%%%%

Acting once again with the operator $\mfD$ on \eqref{GYMeq1} we obtain theintegrability condition
\be
[\mfF\,,\, \cast {\mfF^+}]= 0\,.\label{GYMeq2}
\ee
Note that more generally, the system of equations 
\be\label{universal}
\mfD\mathcal{B}=0\,,\qquad [\mfF,\mathcal{B}]=0\,,
\ee
where $\mathcal{B}$ is a generic differential form and $\mfF$ is the curvature for the connection in $\mfD$, is self-consistent by virtue of the Bianchi identity, $\mfD\mfF\equiv 0$. In fact, acting once more with the covariant derivative produces no new constraints on $\mathcal{B}$.
In the same sense, the equations of motion \eqref{GYMeq1} and their integrability conditions \eqref{GYMeq2} are also self-consistent.

It can be verified that all the components of the commutator \eqref{GYMeq2} along the subalgebra 
\be\label{algh}
\mfh=so(3,1)\oplus \Real \oplus u(1) \oplus su(2)\,,
\ee
vanish identically,
\be
[\mfF\,,\, \cast {\mfF^+}]\LOR \equiv 0\,,\qquad [\mfF\,,\, \cast {\mfF^+}]\DIL \equiv 0\,, \qquad [\mfF\,,\, \cast {\mfF^+}]\INT \equiv 0\,. \label{indentity}
\ee 
Hence, the non-trivial components of \eqref{GYMeq2} are along transvections and supercharge generators;
\be
[\mfF\,,\, \cast {\mfF^+}]\equiv [\, \mfF\,,\, \cast {\mfF^+}]\TRA + [\mfF\,,\, \cast {\mfF^+}]\FER \,,\label{lemma}
\ee 
where 
\bea
[\mfF\,,\, \cast {\mfF^+}]\TRA &=&[ \,  \mfF ^-\,,\, \cast {\mfF}^+\, ]+ [\XX\,,\, \cast \XX]  \,,\label{GYMeq2a}\\[5pt]
[\mfF\,,\, \cast {\mfF^+}]\FER
&=&\overline{\QQ}\, \rho(\, \mfF\BOS\,  i\Gamma- \cast {\mfF}^+\, )D\psi+ D\overline{\psi}\, \rho(\, i\Gamma\, \mfF\BOS- \cast {\mfF}^+\, )\QQ \,.\label{GYMeq2b}
\eea
Therefore, \eqref{GYMeq2} is equivalent to the system
\bea
&
\begin{split}
\Big( \mfF^b \,  (\cast \, {\mfF})_b {}^a +\tilde \mfF^a * \mfF^5 \,  + 2 \, D\overline{\psi}\,  i\gamma_5 B^a \, D{\psi} \; \Big)\JJ_a \hspace{4cm}& \\ 
+ \Big( \tilde \mfF^b \, (\cast \, {\mfF})_b {}^a + \mfF^a * \mfF^5 \, - 2 \, D\overline{\psi}\,  i\gamma_5 \tilde B^a \, D{\psi} \; \Big)\tilde \JJ_a = 0\,,&
\label{GYMeq2c}
\end{split}
\\[5pt]
&
\begin{split}
\overline{\QQ}\; \rho\Big( \, \{\,\mfF\DIL+\mfF\INT\,\}\,(i\Gamma- *)\,-i\Gamma \,\mfF^-\Big)D\psi \hspace{3cm}&\,\\
+ D\overline{\psi}\, \rho\Big(\,(i\Gamma- *) \,\{\,\mfF\DIL+\mfF\INT\,\}\,+\,i\Gamma\mfF^- \Big)\; \QQ& = 0\,.
 \end{split}
& 
\label{GYMeq2d}
\eea
Alternatively, the supercharge-valued constraint can be expressed as
\be
\begin{split}
\overline{\QQ}\; \Big( \, \{\; (F-I)\DIL+\;(F-I)\INT\; \} \; (i\g_5- * ) D\psi  +(F^--{I^-})\;i\g_5  D\psi \Big)  \hspace{1cm}&\,\\
+ \Big(D\overline{\psi} \,(i\g_5- *) \,\{\; (F-I)\DIL+\;(F-I)\INT\; \}\,+D\overline{\psi}\,i\g_5 \;(F-I)\INT\;  \Big)\; \QQ& = 0\,.
 \end{split}
\label{GYMeq2e}
\ee

An $su(2,2|2)$ transformation of the connection gauge field ($\delta\mfA= \mfD \lambda$) and its curvature ($\delta\mfF= [\mfF \,,\,\lambda]$), implies that the Lagrangian changes as
\be\label{deltaL}
\delta_{\lambda}\calL = 2\,\str \left(\lambda \,[\mfF\,,\cast  \mfF^+ ] \right) \; + \; \bterm \, 
\ee
From \eqref{indentity} we know that \eqref{deltaL} vanishes identically for $\lambda \in \mfh$, hence \eqref{algh} is a genuine gauge (off-shell) symmetry of the system. As for transvections and supersymmetry, they are \textit{conditional symmetries}, \ie subjected to their dual symmetry constraints \eqref{GYMeq2c} and \eqref{GYMeq2d} respectively.

%%%%%%%%%%%%%%%%%%%%%%%%%%%%%%%%%%%%%%%%%%%%%%%%%
\section{Ground states and effective theories} \label{sec:effe}
%%%%%%%%%%%%%%%%%%%%%%%%%%%%%%%%%%%%%%%%%%%%%%%%%

We have not yet established the relation between the symmetric tensor $g_{\mu\nu}$, used to build the Hodge dual necessary for the Yang-Mills action, and the transvection gauge fields in the $W^-=P+\tilde P$ component of the gauge connection. 

So far, we have assumed, as in Yang-Mills theories, that the symmetric tensor $g_{\mu\nu}$ is a prescribed function, like a fixed parameter of the action, not dynamical field. It is therefore not varied in the computation of field equations and the symmetry transformations of the Lagrangian. In this picture, the expected correspondence of the type $e^a_\mu\sim p^a_\mu$, $e^a_\mu \sim \tilde p\, {}^a_\mu$, so that $g_{\mu\nu}=e_\mu^a e_\nu^b \: \eta_{ab},\:$ should be established \textit{a posteriori}, as part of the solutions around a ground state. 

In order to avoid the emergence of new fields related to the basis-change matrices,
\be
%\tau_{b}{}^a:=
\frac{\delta  p^a  }{\delta e^b}\,,\qquad %\tilde\tau_{b}{}^a:=
\frac{\delta \tilde p^a  }{\delta e^b}\,, \label{epdelts}
\ee
that could spoil Lorentz invariance, they must be proportional to the only available invariant tensor of rank 2, the Kronecker delta. Hence, following \cite{Townsend:1977xw,Alvarez:2020qmy}, we consider a ground state sector in which the transvection fields are chosen as,
\be \label{ep1}
p^a=\a_+ e^a \,,\qquad \tilde p\,{}^a = \a_- e^a\,,
\ee
with constants $\a_\pm$. 

The field equation for the spin connection,
\be\label{tor1}
\epsilon_{cdab}\Big(\mfF^a p^{b} - \tilde \mfF\,{}^a  \tilde{p}\,{}^b\Big)=0\,,
\ee
is an algebraic equation. When the system of equations \eqref{tor1} is non-degenerate,  the Lorentz connection can be solved in terms of the transvection gauge fields $p^a$, $\tilde{p}{}^a$, and the gravitino currents $I{}^a$ and $\tilde I{}^a$.

In the ground state \eqref{ep1}, for non degenerate \eqref{pab}
\be\label{pp}
p^{ab} = (\a_+^2-\a_-^2)\, e^a e^b\,,
\ee
the equation \eqref{tor1} can be reduced to the torsion constraint
\be 
T^a=\frac{\a_+I^a -\a_-\tilde I^a }{\a_+^2-\a_-^2}\,,\qquad T^a:=D_\Omega e^a\,.
\ee
Hence decomposing the spin connection in a torsionless component (such that $D_{\Omega(e)}e^a=0$) and the contorsion, $\Omega=\Omega(e)+K$, we obtain the  solution
\be \label{solution}
\Omega_\nu^{ab}(e)=2 e^{[a|\rho}\partial_{[\nu} e^{|b]}_\rho - e_{c\nu} e^{[a|\mu} e^{|b]\rho} \partial_\mu e_\rho ^c\,,\qquad K_\mu^{ab}=-\frac12 e^{a\nu}e^{b\rho} (T_{\mu\nu\rho}-T_{\nu\rho\mu}+T_{\rho\mu\nu})\,,
\ee
where $T_{\mu\nu\rho}= T_{\mu\nu}^a \,  e_{a\rho}$.

\subsection{$\calN=2$ supergravity ground state}

Imposing Majorana reality conditions on the gravitino fields, and setting $\a_-=0$, the solution \eqref{solution} produces
\be\label{tor2}
\mfF^a = F^a-I^a =0\,,\qquad  \tilde \mfF\,{}^a = \tilde F\,{}^a-\tilde{I}\,{}^a =0\,.
\ee
Using this back in the transvection symmetry constraint \eqref{GYMeq2c}, we are left with
\be \label{susy2}
D\overline{\psi}\,  i\gamma_5 B^a \, D{\psi} =0 \,,\qquad D\overline{\psi}\,  i\gamma_5 \tilde B^a \, D{\psi} =0\,,
\ee
which can be alternatively written as
\be\label{susy3}
 D\overline{\psi} B (i\gamma_5- *)D\psi - (D\overline{\psi}i\gamma_5- * D\overline{\psi}) B D\psi=0\,,\quad\mbox{with}\quad B=B^a,\,\tilde B\,{}^a\,.
\ee 
Now using  \eqref{tor2} in the supersymmetry constraint \eqref{GYMeq2d}, we get
\be\label{susy4}
\begin{split}
 \{\; (F-I)\DIL+\;(F-I)\INT\; \} \; (i\g_5- * ) D\psi &=0\,,\\[5pt]
  (D\overline{\psi} i\g_5- *D\overline{\psi} ) \,\{\; (F-I)\DIL+\;(F-I)\INT\; \} &=0\,.
\end{split}
\ee 
Since \eqref{susy3} and \eqref{susy4} can be factorized by the Rarita-Schwinger equations \eqref{RSequiv}, 
\be\label{RS2}
(i\g_5- * ) D\psi = 0\,,\qquad D\overline{\psi}i\gamma_5- * D\overline{\psi}=0\,,
\ee
the torsion constraints \eqref{tor2} and the Rarita-Schwinger equation \eqref{RS2} provide enough conditions for transvection symmetry  and supersymmetry.

We stress that \eqref{GYMeq2c} and \eqref{GYMeq2d} could be solved by more general methods, which can allow complex gravitino configurations and non-trivial field strengths.

%%%%%%%%%%%%%%%%%%%%%%%%%%%%%%%%%%%%%%%%%%%%%%%%%%
\subsection{Gravitino ground state}
%%%%%%%%%%%%%%%%%%%%%%%%%%%%%%%%%%%%%%%%%%%%%%%%%%

The supersymmetry constraint \eqref{GYMeq2e} can be fulfilled also in the gravitino vacuum configuration
\be 
D\psi=0\,,\qquad D\overline{\psi}=0\,.\label{sconst1}
\ee
Since $D^2\psi=F\psi$ we also need 
\be\label{F0}
F\psi=0\,.
\ee
In particular, the case $F=0$ implies that all the bosonic curvatures (\ref{Florentz}-\ref{FYM}) must vanish: 
\begin{eqnarray}
0&=& R^{ab}(w)+p^ap^b-\tilde{p}\,{}^a\tilde{p}\,{}^b \,,\label{Florentz2}\\[4pt]
0 &=&  D_\Omega p^a-\oh \tilde{p}\,{}^a =D_\Omega \tilde{p}\,{}^a -\oh {p}^a \,, \label{Ftransv2}\\[4pt]
0&=& d\oh+ p^a \tilde{p}_{a} \,,\label{Fdil2}\\[4pt]
0 &=& G^r \,.\label{FYM2}
\end{eqnarray}
From \eqref{Florentz2} solutions interpolating Anti de Sitter and the de Sitter spaces can be achieved with a suitable choice of the parameters $\a_+$ and $\a_-$ in \eqref{ep1}. The flat case, $R^{ab}(w)=0$, occurs for $\a_+^2=\a_-^2$. This case, however, is degenerate since \eqref{pp} vanishes, which is reflected also by the fact that the Einstein Hilbert term drops out from the Lagrangian \eqref{Lbos1}.
Replacing \eqref{ep1} in the torsion-like conditions \eqref{Ftransv2} this yields,
\be\label{trsncnst}
\a_\pm D_\Omega e^a-\a_\mp \oh e^a =0\,,
\ee
which, in the non-degenerate case $\a_+^2\neq \a_-^2$, requires $\oh=0$, $D_\Omega e^a=0$, and spacetime to be of constant curvature,
\be\label{(A)dScrvtr}
R^{ab}(w)+(\a_+^2 -\a_-^2)e^ae^b=0\,.
\ee

%%%%%%%%%%%%%%%%%%%%%%%%%%%%%%%%%%%%%%%%%%%%%%%%%%%%%%
\subsection{Effective Lagrangian}\label{sec:bi}
%%%%%%%%%%%%%%%%%%%%%%%%%%%%%%%%%%%%%%%%%%%%%%%%%%%%%%
With the transvection fields at their ground states \eqref{ep1}, the theory \eqref{Lagrangian} yields the effective Lagrangian\footnote{By effective we simply mean that the theory can be expanded around the ground state  \eqref{ep1}.} take the form
\be\label{Lagrangianeff}
\calL\eff=- \str \; \mfF_\circ \cast \mfF_\circ\,,
\ee
with $\mfF_\circ=d\mfA_\circ+\mfA_\circ^2$ built from the 1-form
\bea
&\mfA_\circ=\frac12 \omega^{ab}\JJ_{ab}+ \a_+ e^a  \, \JJ_{a} +\a_- e^a \, \tilde\JJ_{a} +\oh\, \JJ_5+ U^r \TT_r\,+ \psi_i^\a \, \overline{\QQ}_\a^i-\overline{\psi}_\a^i \,\QQ_i^\a\,.&\label{corresp}
\eea

The field equations for \eqref{Lagrangianeff} are obtained from \eqref{EoMw1}-\eqref{deltapsi2} taking into account the dependence on the vierbein, implicit in \eqref{ep1}, 
\be\label{e-var-L}
\delta_e\calL\eff= \a_+\,\delta e^a \, \frac{\partial  \calL}{\partial p^a}  + \a_-\, \delta e^a \, \frac{\partial \calL}{\partial \tilde p\,{}^a} + \delta e^a \, \frac{\partial \calL}{\partial e^a}\,.
\ee
Here the partial derivatives indicate functional derivative w.r.t. the explicit dependency on the variables $p$ and $\tilde p$. The first two terms on the right hand side of \eqref{e-var-L} are obtained from the sum of the field equations \eqref{deltap} and \eqref{deltapt} multiplied by $\a_+$ and $\a_-$, respectively. The third term is obtained from the Yang-Mills terms, 
\be\label{YMterms}
\calL_\GG = - \, \str\, ( \GG * \GG )\,,\qquad \GG= \mfF\DIL + \mfF\INT\,.
\ee
Hence
\be\label{deltaeGG1}
\delta_e \calL_\GG=  -  \int d^4x \,e \, \delta e^a_\mu \, V_a^\mu\,, 
\ee
where
\be
V_a^\mu:=\str\left(\GG_{\lambda \rho} \GG^{\lambda \rho} e^\mu_a - \frac14 \GG_{\lambda \rho} \GG^{\lambda \mu} e^\rho_a \right)\,.
\ee
Here $e^\mu_a$ is the inverse of the vielbein and we have also introduced  the inverse metric tensor $g^{\mu\nu}=e^\mu_ae^\nu_b\eta^{ab}$ to raise the 2-form indices of $\GG$. 

With a slightly different parametrization of the linear correspondence \eqref{ep1},
\be \label{ep3}
p^a=\alpha (1-\tau) e^a \,,\qquad \tilde p\,{}^a= \alpha \tau e^a\,,  
\ee
so that $W^- =\frac12 \alpha ((1-\tau)-\tau \g_5) \; \slashed{e}$, the Lorentz curvature reads
\be\label{crvtrcnst1}
F^{ab}=R^{ab} (w) +\a^2(1-2\tau) e^a e^b \,.
\ee
Thus, $\tau$ interpolates between anti de Sitter for $\tau\in (-\infty ,1/2)$ and de Sitter for $\tau\in (1/2,\infty)$, and gives the degenerate case for $\tau=1/2$. The effective Lagrangian \eqref{Lagrangianeff} reads
\bea
&&\begin{split}
\calL\eff_\bos=&  \a^2 \Big( \frac12  -\tau \Big)  R^{ab}(w)e^ce^d\epsilon_{abcd} + \a^4 \left(\frac{1}{2} -\tau\right)^2  e^ae^be^ce^d\epsilon_{abcd}\\[4pt]
&+\frac14 R^{ab}(w)R^{cd}(w)\epsilon_{abcd}  - d\oh * d\oh  -  \frac12 F^I*F^I  -  4dU^6*dU^6 \,, \label{boseff}
\end{split}
\\[7pt]
&&\begin{split}
\calL\eff_\fer&=  4 \overline{\psi} \left[ i \slashed{e} \alpha \left(\frac{ \pi_+}2 - \Big(\frac12-\tau\Big) \pi_-  \right) D_{\Omega +U} + \frac{\a^2}4 \Big(\frac12-\tau\Big) i\g_5 \slashed{e}^2  \right. \\[4pt]
&\hspace{1cm}\left.+ i D\slashed{e} \, \alpha \left(\frac{ \pi_+}2 - \Big(\frac12-\tau\Big) \pi_-  \right)+ \frac12 (* - i \g_5)(F\vert_D+F_U)  - \frac14 \circledast {I^+} \right] \psi\,.
\end{split}\label{fereff}
\eea
In the fermionic sector, $\pi_{\pm}=(\one \pm \g_5)/2$ are the chiral projectors. We observe that the Lagrangian \eqref{fereff} breaks partity (asymmetric chiral terms) and it has a bi-parametric Newton constant. 

Note that the fact that for $\tau=1/2$ the gravity sector in \eqref{boseff} decouples is consistent with the fact that pure Yang-Mills theories provide a good approximate description of internal interactions at short scales, with no need of gravity. 

For positive cosmological constant ($\tau>1/2$) the Lagrangian \eqref{boseff} produces ghosts modes for gravitons, since the Einstein-Hilbert term has the opposite sign. In section \ref{sec:altcast} an alternative Lagrangian is proposed where this is fixed.

%%%%%%%%%%%%%%%%%%%%%%%%%%%%%%%%%%%%%%%%%%%%%%%%%%%%%%%%%%%%%%
\subsubsection{Standard normalization of the Lagrangian}
%%%%%%%%%%%%%%%%%%%%%%%%%%%%%%%%%%%%%%%%%%%%%%%%%%%%%%%%%%%%%%

The standard Einstein-Hilbert and Yang-Mills Lagrangians, 
\bea
S_{\texttt{EH}} &=&\frac{1}{2\kappa^2}  \int \; \frac12 \Big ( R^{ab}(w) -\frac{ \lambda}{3!} e^a e^b\Big) e^c e^d \epsilon_{abcd}= 
\frac{1}{2\kappa^2}  \int d^4x \, e\, (R-2\lambda) \,\label{EH},\\[4pt]
S_{\texttt{YM}} &=& - \frac{1}{2 g_{\scriptscriptstyle{SU(2)}}^2} \int G^I *  G^I- \frac{1}{2 g_{\scriptscriptstyle{U(1)}}^2} \int  G^6 *  G^6\,, \label{EW}
\eea
are contained in the effective Lagrangians \eqref{boseff}-\eqref{fereff} for $\tau\in(-\infty,1/2)$,
\be
 \frac{1}{\kappa^2} = 2\a^2 (1 - 2\tau) \,,\qquad g_{\scriptscriptstyle{SU(2)}}=1\,,\qquad g_{\scriptscriptstyle{U(1)}}=\frac{1}{2\sqrt{2}}\approx 0.35\,. \,\label{cnstnt}
\ee
The Gravity coupling $\kappa$ is bi-parametric and from the first relation in \eqref{cnstnt} $\alpha$ has  the units of the inverse of the Newton constant $G_N$ since $\kappa^{2}=8\pi G_N$. 

In addition, there is a new abelian term in \eqref{boseff} corresponding to the (non-compact symmetry) dilation gauge field $\oh$,
\be
S_{\texttt{dil}} = - \frac{1}{2 g_{\texttt{D}} ^2 } \int  \;  d\oh * d\oh \,, \label{Sdil}
\ee
therefore $g_{\texttt{D}} =1/ \sqrt{2} $. Hence we have the hierarchy $g_{\scriptscriptstyle{U(1)}} <  g_{\texttt{D}} < g_{\scriptscriptstyle{SU(2)}}$. Note that the dilation gauge field $\oh$ is not minimally coupled, but it has a Pauli coupling to the gravitino.

For $\tau=0$ and with a rescaled gravitino field, the standard Rarita-Schwinger action is contained in \eqref{fereff} in the form, 
\be
S_{\texttt{RS}} = - \frac{1}{\kappa^2}  \int  \overline{\zeta}i \g_5\slashed{e}D\zeta=  \frac{1}{\kappa^2}  \int d^4x \, e \; \overline{\zeta}_\mu \gamma^{\mu\nu\lambda} D_\mu \zeta_\nu\,,\qquad  \zeta=\frac{\psi}{\sqrt{\a}}\, . 
\label{RSact}
\ee
For more general values of $\tau$, from the presence of the chiral projectors, the gravitino field should be decomposed in its chiral sectors, which will therefore appear with different weights. 

%%%%%%%%%%%%%%%%%%%%%%%%%%%%%%%%%%%%%%%%%%%%%%%%%%%%%%%%%%%%%%
\subsubsection{Limiting chiral model}
%%%%%%%%%%%%%%%%%%%%%%%%%%%%%%%%%%%%%%%%%%%%%%%%%%%%%%%%%%%%%%

We can obtain fixed chirality gravitino models from the model \eqref{Lagrangianeff} in the limit,
$\alpha \to 0$, $\tau \to -\infty$ while
\be \label{limit}
\a^2 \Big(\frac12-\tau\Big) = \frac{1}{4\kappa^2}
\ee
is kept fixed. Hence we obtain in the bosonic sector  \eqref{boseff},
\begin{eqnarray}
\calL\eff_\bos&=&  \frac{1}{4\kappa^2} 
  R^{ab}(w)e^ce^d\epsilon_{abcd} + \frac{1}{16\kappa^4} 
  e^ae^be^ce^d\epsilon_{abcd}  \\[4pt]
&& - d\oh * d\oh- \frac12 F^I*F^I - 4dU^6*dU^6 +\frac14 R^{ab}(w)R^{cd}(w)\epsilon_{abcd} \,, \label{badEH}
\end{eqnarray}
whilst for the fermion term \eqref{fereff}, redefining $\zeta=\pi_- \psi/\sqrt{\alpha}$,  we get the Rarita-Schwinger action for a chiral field with a torsion-coupling,
\be
\calL\eff_\fer= + \frac{i}{\kappa^2} 
 \; \overline{\zeta} \left[  \slashed{e} D_{\Omega+U}  - \frac12  D\slashed{e} \right] \zeta\,.\label{Lferlim1}
\ee

%%%%%%%%%%%%%%%%%%%%%%%%%%%%%%%%%%%%%%%%%%%%%%%%%%%%%%%%%%%%%%
\subsection{Alternative $\cast_s$ operator and the de Sitter sign fix}  \label{sec:altcast} 
%%%%%%%%%%%%%%%%%%%%%%%%%%%%%%%%%%%%%%%%%%%%%%%%%%%%%%%%%%%%%%

The wrong sign in the Einstein-Hilbert term in \eqref{boseff} for $\tau\in(1/2, \infty)$ can be fixed by redefining the action of the generalized Hodge dual operator on the Lorentz component of the field strength: $\cast \mfF\vert_L \rightarrow -\cast \mfF\vert_L$. This leads to the wrong sign of the Einstein-Hilbert term in the anti de Sitter sector. The Pauli-like term $\overline{\psi}F\vert_L\psi$ produced by the terms $\str ( \FF \LOR \cast \II \LOR )$  will  not cancel the identical term produced by the fermion sector $\str \XX \cast \XX$. Hence, in order to prevent the new Lorentz-Pauli coupling we need to flip also the sign of the $\cast$ operator on the fermionic curvatures: $ \cast \XX\rightarrow - \cast \XX$. 

Different choices for the operator $\cast$ can be selected by introducing the \textit{ad hoc} sign function, 
\bea
s_\tau=
\left\{ 
\begin{array}{l}
\:\:\;1\,,\quad \tau \in (-\infty,1/2]\\
-1\,,\quad \tau \in (1/2, \infty)
\end{array}\right.
\,,
\eea
such that, 
\begin{eqnarray}\label{twtcast}
\cast_s\mfF \vert_L=s_\tau (\cast  \mfF\vert_L) \,,\:\:  \cast_s  \XX =  s_\tau  (\cast  \XX) \,,\:\:
\cast_s  (\mfF \vert_T+\mfF \vert_D + \mfF \vert_U) =* (\mfF \vert_T+\mfF \vert_D + \mfF \vert_U) \,,
\end{eqnarray}
which produces the alternative Lagrangian,
\be\label{Lagrangianalt}
\calL^{\texttt{alt}}:=-  \str \, \mfF^+ \, \cast_s\,  {\mfF }^+\,,
\ee
suitable for both, negative and positive curvature backgrounds.
Since $ (\tau-1/2) \, s_\tau=|\tau-1/2|$,
the new bosonic and fermionic components of the Lagrangian $\calL^{\texttt{alt}}= \calL_\bos^{\texttt{alt}}+ \calL_\fer ^{\texttt{alt}}\,$ are given respectively by,
\be
\begin{split}
\calL^{\texttt{alt}}_\bos=&  \a^2 \Big\vert \frac12  -\tau \Big\vert  R^{ab}(w)e^ce^d\epsilon_{abcd} + s_\tau\, \a^4 \left(\frac{1}{2} -\tau\right)^2  e^ae^be^ce^d\epsilon_{abcd} \\[4pt]
&+ s_\tau\,\frac14 R^{ab}(w)R^{cd}(w)\epsilon_{abcd}  - d\oh*d\oh  -  \frac12 F^I*F^I  -  4dU^6*dU^6\,,\label{bosalt}
\end{split}
\ee
and 
\be
\begin{split}
\calL\eff_\fer=& 4 \overline{\psi} \left[ s_\tau \,i \slashed{e} \alpha \left(\frac{ \pi_+}2 - \Big(\frac12-\tau\Big) \pi_-  \right) D_{\Omega+U} + \frac{\a^2}4 \Big| \frac12-\tau\Big| i\g_5 \slashed{e}^2 \right. \\[4pt]
 &\left.+ s_\tau \,i D\slashed{e} \, \alpha \left(\frac{ \pi_+}2 - \Big(\frac12-\tau\Big) \pi_-  \right)+ \frac12 (* - s_\tau \, i \g_5)(F\vert_D+F_U)  - \frac14 \circledast_s {I^+} \right] \psi\,.
\end{split}
\label{feralt}
\ee
Now  the Einstein-Hilbert term sign is always correct and only the cosmological term in the gravity side flips sign.

The chiral models are obtained as before in the limits $\alpha\rightarrow0,\; \tau \rightarrow \pm \infty$, while keeping fixed the value \eqref{limit}. From \eqref{bosalt} this yields
\begin{eqnarray}
\calL^{\texttt{alt}}_\bos&=& \frac{1}{4\kappa^2} 
 R^{ab}(w)e^ce^d\epsilon_{abcd} + \frac{s_{\pm \infty}}{16\kappa^4} 
 e^ae^be^ce^d\epsilon_{abcd}  \\[4pt]
&&+ \frac{s_{\pm \infty}}4 R^{ab}(w)R^{cd}(w)\epsilon_{abcd}   - d\oh * d\oh -  \frac12 F^I*F^I -  4dU^6*dU^6\,, \label{badEH1}
\end{eqnarray}
where $s_{\pm \infty}=\mp 1$, and in the fermion sector \eqref{feralt}, redefining $\zeta=\pi_- \psi/\sqrt{\alpha}$ gives
\be\label{Lferlim2}
\calL^{\texttt{alt}}_\fer= \frac{i}{\kappa^2}\; \overline{\zeta} \left[  \slashed{e} D_{\Omega+U}  - \frac12  D\slashed{e} \right] \zeta\,.
\ee

%%%%%%%%%%%%%%%%%%%%%%%%%%%%%%%%%%%%%%%%%%%%%%%%%%%%%%%%%%%%%%
\subsection{The matter ansatz} \label{sec:mansatz}
%%%%%%%%%%%%%%%%%%%%%%%%%%%%%%%%%%%%%%%%%%%%%%%%%%%%%%%%%%%%%%

In \cite{Alvarez:2011gd} a mechanism to incorporate Dirac fermion (0-forms)  in a supersymmetric gauge connection was introduced, such that the corresponding 3D Chern-Simons supergravity action produced, instead of a Rarita-Schwinger term, the Dirac action minimally coupled to a Maxwell gauge field and gravity, with a torsion-dependent mass. This approach, referred to as \textit{unconventional supersymmetry}, has been used to build several models in 3D \cite{Alvarez:2015bva,Andrianopoli:2018ymh}, including interesting applications in condensed matter systems \cite{Andrianopoli:2019sip,Iorio:2020tuc,Iorio:2020olc,Gallerati:2021htm,Gallerati:2021rtp}. Extensions of these ideas to four dimensions can be found in  \cite{Alvarez:2013tga,Alvarez:2020qmy}. 

In unconventional supersymmetry (for a review see \cite{Alvarez:2021zhh}) a spin-$1/2$ field is introduced directly in the supersymmetry gauge connection, not as a fundamental gravitino field but combined with the vielbein in the form,
\be \label{mans}
\bfPsi:=\overline{\QQ} \, (\slashed{e} \xi) \,, \qquad \overline{\bfPsi} = \overline{(\slashed{e} \xi)}\,\QQ\, \,,
\ee 
where $\xi$ is a fermion 0-form. Hence, instead the action principle for a spin-$3/2$ field the results is a spin-$1/2$ action principle. This justified to denote \eqref{mans} as the \textit{matter ansatz}.

In reference \cite{Alvarez:2020qmy} an unconventional supersymmetry model was proposed in four dimensions based in the superalgebra $su(2,2|2)$ and a Lagrangian of the type \eqref{Lagrangian}, with the fermion sector replaced by the matter ansatz \eqref{mans}. Similarly, here when the matter ansatz \eqref{mans} is used in \eqref{Lagrangian}, one obtains the descendant Lagrangian
\be\label{Lans}
\calL_\texttt{m-ans} := -\; \str\Big(\mfF_\texttt{m-ans} \circledast \mfF_\texttt{m-ans} \Big)\,.
\ee
In this Lagrangian the fermion field strength \eqref{auxF} is given by $\calX:=\overline{\QQ}_\a^i  \, D(\slashed{e} \xi)^\a_i $. Hence one would obtain a theory for $U(1)\times SU(2)$--minimally coupled matter fermions governed by a Dirac action, with additional torsion and Pauli couplings, and four-fermion self-interactions.

As in \cite{Alvarez:2020qmy}, it can be shown that the coupling constants $g {}_{\scriptscriptstyle{U(1)}}$ and $g {}_{\scriptscriptstyle{SU(2)}}$ for the matter fields $\xi$ respect the electroweak hierarchy $g {}_{\scriptscriptstyle{U(1)}}< g {}_{\scriptscriptstyle{SU(2)}}$, however their values \eqref{cnstnt} are somewhat higher than those of the standard model ($g {}_{\scriptscriptstyle{U(1)}}^{\texttt{SM}}\approx 0.34$, $g {}_{\scriptscriptstyle{SU(2)}}^{\texttt{SM}} \approx 0.66$ \cite{Langacker}). Moreover the Weinberg angle does not correspond to the standard model value.

The model \eqref{Lans} would differ from that in \cite{Alvarez:2020qmy} because the operator $\cast$ used there acts on the fermion component of the curvature with the opposite sign. As a consequence, in \cite{Alvarez:2020qmy} there is an additional coupling of the fermion field and the Lorentz curvature with respect to the one here. 

Without further additions, the theory obtained in this way hinges on the identification between transvection and vierbein fields. Hence, considering the ground states \eqref{ep1}, we can obtain a theory of fermions coupled to gravity and gauge fields in a more standard fashion applying the matter ansatz \eqref{mans} in \eqref{fereff} and \eqref{feralt}, or in the chiral-model limits \eqref{Lferlim1} and \eqref{Lferlim2}. 
In particular a matter--anti-matter symmetry breaking of  fermions can be fine tuned using the parameter $\tau$ in \eqref{fereff} and \eqref{feralt}. Instead in \eqref{Lferlim1}-\eqref{Lferlim2} the chirality of the fermions is fixed. See \cite{Alvarez:2020qmy} for further discussions.

%%%%%%%%%%%%%%%%%%%%%%%%%%%%%%%%%%%%%%%%%%%%%%%%%
\section{Conclusions}\label{sec:conc}
%%%%%%%%%%%%%%%%%%%%%%%%%%%%%%%%%%%%%%%%%%%%%%%%%

We have constructed a $\calN=2$ supergravity model based on the $G=SU(2,2|2)$ symmetry, whose precise correspondence with the usual $\mathcal{N}=2$ supergravities 
\cite{deWit:1979dzm,deWit:1980lyi,deWit:1983xhu,deWit:1984wbb,deWit:1984rvr,Castellani:1990tp,DAuria:1990qxt} is far from obvious. The %field contents 
gauge symmetry of those models and ours is similar, but there are several important differences as well: We do not include matter fields, all our fields come from the gauge connection. We treat the gauge fields $p^a$ and $\tilde{p}^a$, associated to the transvection sector of the superalgebra, on equal footing, whilst in the previous approaches the second frame-like field is resolved in terms of the remaining fields, invoking certain additional constraints (see \eg \cite{deWit:1979dzm,deWit:1980lyi}). Instead of two types of gravitino fields, our gravitino is complex and charged with respect to $U(1)\times SU(2)$ interactions governed by the standard Yang-Mills theory.

Our approach can be construed in the context of the group theoretical approach of supergravity \cite{Neeman:1978zvv,DAdda:1980axn,Castellani:1981um}. Here, the transvection symmetry and supersymmetry are both broken off-shell, but they can be regarded as conditional symmetries valid if the integrability conditions of the field equations hold. 

The ground state \eqref{ep1}, in which the transvection fields are proportional to the frame fields generically corresponds to (anti-)de Sitter vacuum. The effective theory around this vacuum contains $\calN=2$ pure supergravity configurations, with Majorana fermions satisfying the Rarita-Schwinger equation interacting with standard Einstein-Hilbert gravity, $U(1)\times SU(2)$ gauge fields and the frame field satisfying the regular torsion constraint.

In the bi-parametric class of effective models obtained around those ground states, the parameter $\tau$ determines simultaneously the chiralities of the gravitino, the gravitational coupling constant and the cosmological constant. Therefore this parameter controls the balance of the matter---anti-matter modes, simultaneously with the gravity coupling and the cosmological constant. In the degenerate limit $\tau=1/2$, the gravity Lagrangian drops out and only the Yang-Mills terms together with a chiral Rarita-Schwinger action remain. These features are inherited by the model of spin-$1/2$ Dirac (chiral) field obtained by means of the matter ansatz discussed in section  \ref{sec:mansatz} (\cf \cite{Alvarez:2020qmy}). Hence, this supergravity theory can produce models of a realistic sort in two steps: firstly expanding around the transvection ground states \eqref{ep1}, secondly projecting to the spin-$1/2$ component of the gravitino by means of the matter ansatz.

The existence of two types of frame fields, $p^a$ and $\tilde{p}\,{}^a$ and the breaking of (off-shell) transvection symmetry and supersymmetry, suggests the existence of a larger framework where a \textit{spontaneous symmetry breaking} mechanism would fix the values of the transvection fields. References \cite{Stelle:1979aj,Stelle:1979va,McCarthy:1985nt,Wilczek:1998ea,Magueijo:2013yya,Zlosnik:2016fit} have explored different ways of implementing possible scenarios and may serve as an inspiration in the search of such a mechanism.

Our construction can be extended to other gauge supergroups such as $SU(2,2|N)$ or $OSp(4|N)$, where the same pattern of field equations, integrability conditions equivalent to symmetry constraints, and conditional symmetries can be seen in the sequence:
\be
\delta \int \str \mfF \cast \mfF =0  \;\;\overset{\hbox{EoM}}{\longrightarrow}  \;\; \mfD \cast {\mfF^+} =0 \;\;\;\overset{\hbox{Integrability}}{\longrightarrow}\;\;\; \mfD^2 \cast {\mfF^+}=[ \mfF\,, \cast {\mfF^+}]=0\,.\nonumber
\ee
We leave these extensions as well as those to higher dimensions, the search of solutions and the analysis of quantum aspects, for future works.  

\section*{Acknowledgements}

We thank Per Sundell for useful discussions. This work has been partially funded by Fondecyt grant 1180368, by MINEDUC-UA project ANT1755, and by Semillero de Investigaci\'on project SEM18-02 from Universidad de Antofagasta, Chile.

\appendix
%%%%%%%%%%%%%%%%%%%%
\section{Representation of $su(2,2|2)$}\label{app:su222}
%%%%%%%%%%%%%%%%%%%%

For the sake of definiteness, here we use the following representation for the Gamma matrices:
\bea
 \g^a=
\left(
\begin{array}{cc}
0 & \sigma^a \\
\bar{\sigma}^a & 0
\end{array}
\right)\,, \quad \pi_\pm:=\frac{\mathds{1}\pm\gamma_5}{2}\,, \quad \g_5:=i\gamma^0 \gamma^1 \gamma^2 \gamma^3=\left( 
\begin{array}{cc}
-\one & 0\\
0& \one
\end{array}
\right)\,,\label{Weyl}
\eea
where $\sigma^a= \bar\sigma_a=\{\mathds{1}\, ,\vec{\sigma} \}$ are the Pauli matrices and $\{\gamma^a,\gamma^b\}=2 \, \eta^{ab}$, with Minkowski metric in the signature $\eta=\mathrm{diag}(-,+,+,+)$.

The representation of $su(2,2\vert2)$ superalgebra used in this article is constructed as follows (for more details see  \cite{Kac:1977qb}\cite{Parker:1980af}). The $sl(4\vert 2, \mathds{C})$ is the set of matrices of $M_{4 \vert 2}(\mathds{C})$ with vanishing supertrace and $su(2,2,\vert2)$ is defined as the real form of $\mathfrak{sl}(4\vert 2, \mathds{C})$,
\be
\phi(X) = X  \,,\qquad \phi : X \mapsto - \mathcal{A}^{-1} X^{\dagger} \mathcal{A} \label{realformcond} %\label{realform}
\ee
where $\phi$ is associated to the real structure  with
\be
\mathcal{A} = \left[\begin{array}{c|c}
 i\gamma^0 &  0_{4\times2}\\[0.5em] \hline
0_{2\times4} & i\sigma^0 \\
\end{array}\right]\,, 
\ee
Using this definition, one obtains the following basis:
\begin{eqnarray}
&\JJ_{ab} = \left[\begin{array}{c|c}
 \frac12 \g_{ab} &  0_{4\times2}\\ \hline
0_{2\times4} & 0_{2\times2} \\
\end{array}\right]
\,,\qquad
\JJ_5 = \left[\begin{array}{c|c}
\frac12 \g_5 &  0_{4\times2}\\ \hline
0_{2\times4} & 0_{2\times2} \\
\end{array}\right],& \\
&
\JJ_{\,a}  = \left[\begin{array}{c|c}
\frac12 \g_{a}  &  0_{4\times2}\\  \hline
0_{2\times4} & 0_{2\times2} \\
\end{array}\right]\,,\qquad
\tilde{\JJ}_{a} = \left[\begin{array}{c|c}
\frac12 \g_{a}\g_5  &  0_{4\times2}\\ \hline
0_{2\times4} & 0_{2\times2} \\
\end{array}\right] 
\,, &\\
&\TT_{6}=  \left[\begin{array}{c|c}
i\one_{4\times 4} &0_{4\times2}\\   \hline
0_{2\times4} & 2i\one_{2\times 2}\end{array}\right]\,,\qquad
\TT_{I=7,6,9} = \left[\begin{array}{c|c}
0_{4\times4} &  0_{4\times2}\\ \hline
0_{2\times4} & -\frac12 i (\sigma_{I})^t \\
\end{array}\right]\,,&\\
&(\mathds{O}_1)_i^\alpha = \left[\begin{array}{c|c}
 0_{4 \times 4} &  E_{\alpha i}\\ \hline
-E_{i \beta} (\gamma_0)^\alpha{}_\beta & 0_{2 \times 2} \\
\end{array}\right]\,, \qquad 
(\mathds{O}_2)_i^\alpha = \left[\begin{array}{c|c}
 0_{4 \times 4} &  iE_{\alpha i}\\ \hline
 i E_{i \beta} (\gamma_0)^\alpha{}_\beta & 0_{2 \times 2} \\
\end{array}\right]&
\end{eqnarray}
where we $\sigma$'s are relabeled Pauli matrices,
\be \label{pauli}
\sigma_7:=\sigma_1\,,\qquad  \sigma_8:=\sigma_2\,,\qquad \sigma_9:=\sigma_3\,,
\ee
and $E_{i \alpha}$ denotes the elementary matrix having a non vanishing component $1$ only in the $(i,\alpha)$ entry. 

The $\mathds{O}_1$ and $\mathds{O}_2$ generators are not preserved by the adjoint action of the $su(2)$ generators, they do not span  $su(2)$ singlets.  This is why we use instead the generators:
\be
\mathds{Q}^\alpha_i = \left[\begin{array}{c|c}
0_{4\times4} & 0_{4\times 2}\\   \hline
E_{i \alpha} & 0_{2\times2}
\end{array}\right]\,,\qquad
\overline{\mathds{Q}}_\alpha^i= \left[\begin{array}{c|c}
0_{4\times4} & E_{\alpha i} \\ \hline
0_{2\times 4} & 0_{2\times2}
\end{array}\right]\,.
\ee
Now these generators do not satisfy \ref{realformcond}, so they do not belong to (Real form) $su(2,2|2)$. However, one can show that if the gauge fields coefficients of $\QQ$ and $\overline{\QQ}$ are Dirac conjugate of one another, then the corresponding difference is part of the algebra:
\be 
\overline{\QQ}\psi - \overline{\psi} \QQ\; \;\in\;\; su(2,2|2)\,.
\ee 
Hence the gauge connection $\mfA$ \eqref{fkA} belongs to $su(2,2|2)$ since it satisfies the reality condition \eqref{realformcond}.

The commutation relation of the bosonic subalgebra read:
\begin{equation}
[\mathds{J}_{ab},\mathds{J}_{cd}]=-\eta_{ac}\mathds{J}_{bd}+\eta_{ad}\mathds{J}_{bc}+\eta_{bc}\mathds{J}_{ad}-\eta_{bd}\mathds{J}_{ac}\,. \label{BB1}
\end{equation}
\begin{equation}
[\JJ_{ a}, \JJ_5 ]= \tilde{\JJ}_{a}\,,\qquad [\tilde\JJ_{ a}, \JJ_5 ]= \JJ_{a}\, \label{BB2}
\end{equation}
\begin{equation}
[\JJ_{a},\JJ_{bc}]=\eta_{ab}\JJ_{c}-\eta_{ac}\JJ_{ b}\,,\quad [\tilde\JJ_{a},\JJ_{bc}]=\eta_{ab}\tilde\JJ_{c}-\eta_{ac}\tilde\JJ_{ b}\,,\quad [\JJ_{a},\tilde\JJ_{b}]=\eta_{ab}\JJ_5 \,, \label{BB3}
\end{equation}
\begin{equation}
 [\TT_I,\TT_J]=\epsilon_{IJ}^{\ \ \ K}\TT_K\,,\qquad I=1,2,3\,. \label{BB4}
\end{equation}
The commutation of bosonic and fermionic generators, $[\BB\,,\,\QQ]$ and $[\overline{\QQ}\,,\,\BB]$, are given in \eqref{com} with constants of structures \eqref{rhoJJ} and \eqref{rhoTT}.

The anti-commutation relation of read,
\be
[\mathds{Q}^\alpha_i,\overline{\mathds{Q}}^j_\beta]_+=\begin{array}{c}
\Big((B^{a})^{\alpha\, j}_{i\, \beta} \JJ_{a} -(\tilde B^{a})^{\alpha\, j}_{i\, \beta}  \tilde\JJ_{a} -\frac12(B^{ab})^{\alpha\, j}_{i\, \beta}  \JJ_{ab}+(B_{5})^{\alpha\, j}_{i\, \beta}  \JJ_5 \Big) \\[4pt]
+\Big( 2 (B_I)^{\alpha\, j}_{i\, \beta}  \TT_I +\frac14 (B_6)^{\alpha\, j}_{i\, \beta}  \TT_6 \Big)\,.
\end{array}\label{anticom0}
\ee

The supertrace $\str$ is defined in term of the trace $\tr$ of block-matrices as,
\be
\str\, \mathds{M} = \tr A_{4\times 4}- \tr D_{2\times 2}
 \,, \qquad \mathds{M} = \begin{pmatrix}
A_{4\times 4} &|& B_{4\times 2} \\[-8pt]
-- & &--\\[-8pt]
C_{2\times 4} &|&D_{2\times 2}
\end{pmatrix} \label{Str}
\,,
\ee
which provides the Killing metric elements,
\bea
\begin{gathered}
\str\,( \JJ_{ab} \JJ_{cd}) = 2(\eta_{ad} \eta_{bc}-\eta_{ac} \eta_{bd}) \,,\qquad \str\,( \JJ_a \JJ_b )= -\str \,(\tilde\JJ_a \tilde\JJ_b)=\eta_{ab}  \\[5pt]
\str \,(\JJ_5 \JJ_5)=1\,,\qquad \str \,(\TT_6 \TT_6 )= 4 \,,\qquad \str \,(\TT_I \TT_J )= \frac12 \delta_{IJ}\,,\\[4pt]
\str \,(\overline{\QQ}_\b^j \, \QQ_i^\a )=-\str \,(\QQ_i^\a\,  \overline{\QQ}_\b^j )=\delta_i^j\delta^\a_\b\,.
\end{gathered}
\eea
The inverse Killing form obtained from here, can be also obtained from the commutation relation \eqref{anticom0} and \eqref{anticom1}. 
The fact that the anticommutation of the supercharges can be written using the inverse Killing form is a peculiar property of the algebras $su(2,2|N)$.

\section{Construction the Lagrangian and the $\cast$ operator}\label{sec:Lconst}

In order to construct the Lagrangian \eqref{Lagrangian}  we require that the choice of the $\cast$ should be such that:
\begin{itemize}
\item[i)] $\str\; (\FF\FER\circledast \FF\FER)$ contains  Rarita-Schwinger terms.
 
\item[ii)] $\str\; (\FF\ST \circledast \FF\ST )$ contains Einstein-Hilbert terms.

\item[iii)] $\str\; (\FF\INT\circledast \FF\INT)$ contains the Yang-Mills term.

\item[iv)] The action does not contain torsion kinetic terms.

\end{itemize}

Inspecting these terms we observe that the goal is achieved with the following actions of the generalized Hodge operator, specified on the different sectors of the field strength:
\begin{itemize}
\item[1)] $\circledast \XX= i\Gamma\XX + \XX i\Gamma = \overline{\QQ}i\g_5\calX -\overline{\calX }i\g_5\QQ$. 
 
\item[2)] $ \circledast \FF\LOR= i\Gamma \FF\LOR$\,.

\item[3)] $\circledast \FF\INT=*\, \FF\INT$.

\item[4)] $ \circledast \FF\TRA= i\Gamma  \FF\TRA=-i \tilde F^a \JJ_a-i  F^a \tilde \JJ_a$. Here, even though $\cast$ produces imaginary factors, in the Lagrangian these terms will cancel out.

\item[5)] In addition we choose $ \circledast \FF\DIL= *\, \FF\DIL$ upon the dilation sector. The option $ \circledast \FF\DIL= i\Gamma \FF\DIL$ does not belong to the algebra hence it is discarded. The option $ \circledast \FF\DIL = i\FF\DIL$ yields an imaginary term in the Lagrangian, hence we avoid it. 
\end{itemize}
The $\cast$ operator will act in the same way on any $\mfg$-valued 2-form, in agreement with their $su(2,2|2)$. 

The requirements above are satisfied by the choice \eqref{castdef}, which with more details read, 
\bea
 \circledast \mfF &=&\frac14  ( F^{ab}-I^{ab}) \epsilon_{abcd}\JJ^{cd} -i (\tilde F^a-\tilde{I}^{a}) \JJ_{a} -i (F^a-I^{a})  \tilde\JJ_{a} + \ast (F^5- I^5) \JJ_5 \nonumber \\
&& + * ( G^r-I^r) \TT_r + i \overline{\QQ} \gamma_5 \calX - i \overline{\calX} \gamma_5 \QQ \,. 
\eea
The option used in \eqref{twtcast} also fulfills the requirements.

Since the $\cast$ operator removes the transvection type of terms from the Lagrangian, \eqref{Lagrangian} can be alternatively written as,
\be \label{Lagrangianalt2}
\begin{split}
\calL &=
- \str \; \mfF^+  \circledast \mfF^+ \\ 
&=- \str \; (\FF^+-\II^+)  \circledast (\FF^+-\II^+) - \str \; \XX \circledast \XX\,.
\end{split}
\ee
Here we can see why the imaginary components of $\cast$, on transvections, do not produce imaginary terms in the Lagrangian.

Since the supertraces of the bosonic generators \eqref{strB}  produce the Killing form of the bosonic subalgebra, and since  $\circledast \FF$ is in the algebra by construction (see \eqref{castdef}), the bosonic Lagrangian is equivalent to,
 \be
\calL_\bos:=- (F^+)_N  (\circledast F^+)^N  \,,
\ee
where $(F^+)_N:=(F^+)^M\mathcal{K}_{MN}$.  For the fermionic component we get,
\be
\calL_\fer=  4 \overline{\psi} \left( i\g_5 W^-D  + \frac12\left( \circledast F^+ -  i \g_5 F\right) - \frac14 \circledast {I^+} \right) \psi- d\,(\,2i\overline{\psi}\g_5D\psi\,)\,.\label{Lferm1}
\ee
Noticing that $i\g_5F=i\g_5F^++i\g_5F^-$ and that $\cast F^+=i\g_5 F\LOR + * F\DIL+*F\INT$,  the Lorentz components of the Pauli terms in \eqref{Lferm1}  cancel out,
\be
\circledast F^+ -  i \g_5 F=(\circledast - i \g_5) F^+-  i \g_5 F^-=(* - i \g_5)(F\DIL+F\INT)-  i \g_5 F^-\,.
\ee
Thus we can write $\calL_\fer$ as in \eqref{Lfer1}.

%%%%%%%%%%%%%%%%%%%%%
\section{Forms of the Rarita-Schwinger equation}\label{app:RS}
%%%%%%%%%%%%%%%%%%%%%

Here we follow reference \cite{Deser:1977ur}. Let $B$ be a spinor 2-form satisfying,
\be
\slashed{e}\wedge \Omega=0\,.\label{RRS}
\ee
Then it follows:
\bea
\gamma^{\mu}*\Omega_{\mu\nu}&=& 0\,,\qquad *\Omega_{\mu\nu}:=\frac12 e\, \epsilon_{\mu\nu\lambda\rho}\Omega^{\lambda\rho}\,,\label{RRS1}\\
\gamma^{\mu\nu\lambda}\Omega_{\nu\lambda}&=&0\,, \label{RRS2} \\
\gamma^{\mu}\Omega_{\mu\nu}&=&0\,,\label{RRS3} \\
(i\gamma_5- *)\Omega&=&0\,.\label{SDB}
\eea

From \eqref{RRS}, equivalent to $\gamma_{[\mu} \Omega_{\nu\lambda]}=0$, we demonstrate these  identities performing the following operations:
\begin{itemize}
\item $\epsilon_{\rho\mu\nu\lambda} \gamma^{[\mu} \Omega^{\nu\lambda]}=0$  $\quad \Rightarrow \quad$ 	\eqref{RRS1}
\item from \eqref{RRS1} using identity $i\gamma_5\gamma_{\mu\nu\lambda}=- e\, \epsilon_{\mu\nu\lambda\rho}\gamma^\rho$ $\quad \Rightarrow \quad $  \eqref{RRS2}
\item $\gamma^\mu \gamma_{[\mu} \Omega_{\nu\lambda]} =0 $ and \eqref{RRS2} $\quad \Rightarrow  \quad  $ \eqref{RRS3}
\item we multiply  \eqref{RRS1} and \eqref{RRS3} by $\gamma_\lambda$ and  $i\gamma_5\gamma_\lambda$  respectively, then we add the both terms and anti-symmetrize the 2 free indices to obtain,
\be
\gamma_{[\lambda|} \gamma^{\mu}\Omega_{\mu|\nu]} + i\gamma_5\gamma_{[\lambda|}\gamma^{\mu}*\Omega_{\mu|\nu]}=\Omega_{\lambda\nu} + i\gamma_5 * \Omega_{\lambda\nu}=0\,,
\ee
\end{itemize}
which is equivalent to \eqref{SDB}. In particular these results are valid for $\Omega=D\psi$, hence,
\be\label{RSequiv}
\slashed{e}D\psi=0 \quad \cong\quad  (i\gamma_5- *)D\psi=0\,.
\ee

\end{document}